\documentclass[lettersize,journal]{IEEEtran}
\usepackage{amsmath,amsfonts}
\usepackage{algorithmic}
\usepackage{algorithm}
\usepackage{array}
\usepackage[caption=false,font=normalsize,labelfont=sf,textfont=sf]{subfig}
\usepackage{textcomp}
\usepackage{stfloats}
\usepackage{url}
\usepackage{verbatim}
\usepackage{graphicx}
\usepackage{cite}

\newtheorem{theorem}{Theorem}[section]
\newtheorem{lemma}{Lemma}[section]

\newtheorem{definition}{Definition}[section]

\begin{document}

\title{Privacy-Preserving Polynomial Computing Over Distributed Data}

\author{Zhiquan Tan$^*$, Dingli Yuan$^*$, Zhongyi Huang$^*$\\
$*$ Department of Mathematical Sciences, Tsinghua University, Beijing, China \\}

\maketitle

\begin{abstract}
In this letter, we delve into a scenario where a user aims to compute polynomial functions using their own data as well as data obtained from distributed sources. To accomplish this, the user enlists the assistance of $N$ distributed workers, thereby defining a problem we refer to as privacy-preserving polynomial computing over distributed data. To address this challenge, we propose an approach founded upon Lagrange encoding. Our method not only possesses the ability to withstand the presence of stragglers and byzantine workers but also ensures the preservation of security. Specifically, even if a coalition of $X$ workers collude, they are unable to acquire any knowledge pertaining to the data originating from the distributed sources or the user.
\end{abstract}


\section{Introduction}

\IEEEPARstart{I}{n} the information age, the size of datasets often grows rapidly, rendering their management infeasible using a single server. Consequently, data is frequently distributed across multiple servers that operate in parallel \cite{shvachko2010hadoop}. While distributing computations across multiple servers offers numerous advantages, it also introduces new complexities and challenges. One of the primary challenges is the presence of stragglers, denoting workers that exhibit significantly slower response times than their counterparts \cite{huang2017gray}. This can lead to delays in overall computation and negatively impact system performance. Another concern arises from the existence of malicious workers, commonly referred to as byzantine workers, who may deliberately submit adversarial results for personal gain, thereby jeopardizing the integrity and accuracy of computations. Data privacy is also a significant concern, as certain workers may collude to gain access to sensitive processed data. Addressing these challenges necessitates the development of robust and efficient distributed algorithms.

As previously emphasized, datasets are frequently distributed. In our study, we examine a scenario in which a user possesses private data and seeks to compute a (polynomial) function involving their own data as well as data stored in distributed sources. We termed this problem as privacy-preserving polynomial computing over distributed data. Our objective is to devise a protocol that offers the following features:

\begin{itemize}
\item Resilience against the presence of straggling workers.
\item Robustness against byzantine workers.
\item (Information-theoretic) privacy of sources and user data, even in the event of collusion among workers.
\end{itemize}

In recent years, there has been a surge of interest in integrating coding-theoretic methods \cite{lee2017speeding} into the design of distributed algorithms that exhibit resilience against straggling and byzantine workers while also ensuring data privacy. These methods have proven effective in addressing the challenges associated with large-scale distributed computations. For example, works such as \cite{yu2020straggler, dutta2019optimal, tang2018erasure} propose coded matrix designs to mitigate the impact of stragglers in distributed matrix multiplication. Furthermore, studies like \cite{kim2020fully, zhu2020secure} consider both privacy and the effects of straggling in distributed matrix multiplication. For general distributed polynomial computing problems, Lagrange coded computing (LCC) \cite{yu2019lagrange} provides a scheme that resists the influence of stragglers and byzantine workers, while also ensuring data privacy even in the presence of colluding workers.

In this letter, we propose an approach based on Lagrange encoding to address the challenges posed by privacy-preserving polynomial computing over distributed data. Our proposed method is specifically designed to be resilient against both straggling and byzantine workers. Additionally, it guarantees data security by preventing any coalition of $X$ colluding workers from accessing information pertaining to the data from distributed sources and the user.

\textbf{Notation}: We denote the set of integers from $1$ to $L$ as $[L]$.

\section{Problem Setting}
Assume all the computation shall be performed on a given finite field $\mathbb{F}_q$. We shall consider a scenario where there are $S$ sources and each source $i$ holds some secret data $W_i \in \mathbb{F}^{a \times b}_{q}$. Denote the data jointly shared by these sources as $W \in \mathbb{F}^{a \times bS}_{q}=[W_1 W_2 \cdots W_S]$. Suppose these data are further divided into $W=[W^{(1)} W^{(2)} \cdots W^{(K)}]$. where we assume $S|K$ for ease of exposition. A master also has some data $U=[U^{(1)} U^{(2)} \cdots U^{(K)}]$, where $U^{(i)} \in F^{a \times b\frac{S}{K}}_{q}$. Then the goal is to compute polynomial functions $h(W^{(i)}, U^{(i)})$ ($1 \leq i \leq K$) with the help of $N$ distributed workers. Sources will not communicate with the user, nor will there be communication among sources. In addition, all the workers are
connected to the user and sources. We use the widely adopted setting in coded computing that all the connected links are error-free \cite{kim2020fully, zhu2020secure}.

We shall consider a communication protocol formulated generally as follows:

$\bullet$ \textbf{Sharing}: 
The sharing operation may consist of two parts:
\begin{enumerate}
    \item Each source $i$ may generate a set of random matrices $P^{(i)}$ and choose a set of functions $\{f^{(i)}_1, f^{(i)}_2, \cdots, f^{(i)}_N \}$ then send each worker $k$ encoded data ${\bar W}^{(i)}_k=f^{(i)}_k(W_i,P^{(i)})$.
    \item The master may generate a set of random matrices $Q$ and choose a set of functions $\{g_1, g_2, \cdots, g_N \}$ then send each worker $k$ encoded data ${\bar U}_k=g_k(U, Q)$.
\end{enumerate}

$\bullet$ \textbf{Computing}: After receiving the encoded matrices ${\bar W}^{(i)}_k$ ($1 \leq i \leq S$) and ${\bar U}_k$, worker $k$ shall calculate a matrix $Y_k$ and return $Y_k$ to the user.

$\bullet$ \textbf{Reconstruction}: After receiving any $M$ responses from workers, the user is able to retrieve $h(W^{(i)}, U^{(i)})$ ($1 \leq i \leq K$). We shall call this number $M$ recovery threshold of this protocol. There are also some system cost metrics that should be taken into account:

\begin{enumerate}
    \item Source Upload Cost: For each source $i$, the upload cost $U_{S_i}$ is defined as $\sum_{k \in [N]} H({\bar W}^{(i)}_k)$.
    \item User Upload Cost: $U_u=\sum_{k \in [N]} H({\bar U}_k)$.
    \item User Download Cost: \begin{equation}
      D = \max _{\mathcal{K}: \mathcal{K} \subseteq[N],|\mathcal{K}|=M} \sum_{k \in \mathcal{K}} H\left(Y_k\right).  
    \end{equation}
\end{enumerate}

We would like to design a protocol under the following constraints:

\begin{enumerate}
    \item Data privacy: The protocol should keep workers (information theoretic) $X$-private about the data stored in sources and user. Specifically, 
\begin{equation}
{I}\left(W, U; \widetilde{W}_{\mathcal{X}}, \widetilde{U}_{\mathcal{X}}\right) = 0    
\end{equation}

, for any $\mathcal{X} \subset[N],X=|\mathcal{X}|$. $\widetilde{W}_{\mathcal{X}} = \left\{ \{{\bar W}^{(i)}_k \}_{i \in [S]} \right\}_{k \in \mathcal{X}}$ denotes all the information received from sources by workers in $\mathcal{X}$, $\widetilde{U}_{\mathcal{X}}$ defines similarly.

    \item Byzantine worker robustness: The user shall get the correct answers $f(W^{(i)}, U^{(i)})$ ($1 \leq i \leq K$) even if any $A$ workers send (arbitrary) erroneous responses. A protocol that guarantees robustness against any $A$ byzantine workers is called $A$-secure.
    \item Straggler resilience: The user shall get the correct answers $f(W^{(i)}, U^{(i)})$ ($1 \leq i \leq K$) even if any $B$ workers fail to respond. A protocol that guarantees resilience against any $B$ stragglers is called $B$-resilience.
\end{enumerate}

\section{A computation strategy based on Lagrange encoding}

\subsection{General description of the proposed method}

We select any $K+T$ distinct numbers $\beta_j \in \mathbb{F}_q$ ($1 \leq j \leq K+T$). $N$ distinct numbers $\alpha_i \in \mathbb{F}_q$ ($1 \leq j \leq N$) are chosen under the requirement $\{ \alpha_i \}_{i \in [N]} \cap \{ \beta_j \}_{j \in [K]} =\emptyset$. Then the encoding polynomials are given as follows:

\begin{align} \label{encode user}
g(z) =  & \sum_{j \in[K]} U^{(j)}  \prod_{l \in[K+X] \backslash\{j\}} \frac{z-\beta_l}{\beta_j-\beta_l}+ \nonumber\\
& \sum_{j=K+1}^{K+X} Q_j  \prod_{l \in[K+X] \backslash\{j\}} \frac{z-\beta_l}{\beta_j-\beta_l}.
\end{align}

\begin{align} \label{encode source}
 & f^{(i)}(z)    = \hspace{-2mm}\sum_{j=K+1}^{K+X} P^{(i)}_j  \prod_{l \in[K+X] \backslash\{j\}} \frac{z-\beta_l}{\beta_j-\beta_l} + \nonumber \\
& \sum_{j \in [\frac{K}{S}]}  W^{((i-1)\frac{K}{S}+j)} \hspace{-6mm}  \prod_{l \in[K+X] \backslash\{(i-1)\frac{K}{S}+j\}} \frac{z-\beta_l}{\beta_{(i-1)\frac{K}{S}+j}-\beta_l}.
\end{align}

Whenever each worker $k$ receives all the encoded matrices ${\bar W}^{(i)}_k=f^{(i)}(\alpha_k)$ and ${\bar U}_k=g(\alpha_k)$ from user and all sources, it shall compute $Y_k=h(\sum^{S}_{i=1}{\bar W}^{(i)}_k, {\bar U}_k)$ and return $Y_k$ to the user.

\begin{theorem}
The proposed computation strategy is $X$-private and $A$-secure. It has a recovery threshold of $M=(K+X-1)deg(h)+2A+1$. As long as $M \leq N-B$, this scheme is also $B$-resilience.
\end{theorem}

\textbf{Proof of theorem $3.1$}:

Note 
\begin{align}
    \sum^{S}_{i=1}{\bar W}^{(i)}_k &= \sum^{S}_{i=1} f^{(i)}(\alpha_k) \nonumber\\
&=\sum_{j=K+1}^{K+X} (\sum^{S}_{i=1} P^{(i)}_j)  \prod_{l \in[K+X] \backslash\{j\}} \frac{\alpha_k-\beta_l}{\beta_j-\beta_l} + \nonumber\\
& \sum_{j \in[K]} W^{(j)} \prod_{l \in[K+T] \backslash\{j\}} \frac{\alpha_k-\beta_l}{\beta_j-\beta_l}.
\end{align}

We shall denote $P_j=\sum^{S}_{i=1} P^{(i)}_j$ ($j \in [K]$). 

Define
\begin{align}
f(z) =  & \sum_{j \in[K]} W^{(j)}  \prod_{l \in[K+X] \backslash\{j\}} \frac{z-\beta_l}{\beta_j-\beta_l}+ \nonumber\\
& \sum_{j=K+1}^{K+X} P_j  \prod_{l \in[K+X] \backslash\{j\}} \frac{z-\beta_l}{\beta_j-\beta_l}.
\end{align}

It is clear that $Y_k=h(f(z),g(z))|_{z=\alpha_k}$. As $f(z)$ and $g(z)$ are both uni-variate polynomial of degree $K+X-1$, thus $h(f(z), g(z))$ is a (uni-variate) polynomial of degree $(K+X-1)deg(h)$. The computational result of each worker $k$ is equivalent to evaluating the $k$-th codeword symbol of a Reed-Solomon (RS) code. The results received from byzantine workers can be seen as an erroneous codeword symbol of this RS code. Thus receiving any $(K+X-1)deg(h)+2A+1$ computational results from workers will be sufficient to recover the polynomial $h(f(z), g(z))$ by RS code decoding algorithm. Note for any $j \in [K]$, $h(W^{(j)}, U^{(j)})=h(f(z),g(z))|_{z=\beta_j}$. The proof of privacy can be found in section \ref{privacy proof}. Thus the conclusion follows.

We shall then provide the complexity analysis of the proposed scheme in the following theorem.

\begin{theorem}

\begin{enumerate}
    \item Source (User) Upload Cost: $U_{S_i}=U_u=Nab\frac{S}{K}$.
    \item User Download Cost: $D=((K+X-1)deg(h)+2A+1)ab\frac{S}{K}$.
    \item Decoding Complexity: $O(ab\frac{S}{K}\hat{M} \log^2 \hat{M} \log \log \hat{M})$, where $\hat{M}=(K+X-1)deg(h)+2A$. 
\end{enumerate}
\textbf{Proof of theorem $3.2$}:
The upload costs follow from equations \ref{encode user} and \ref{encode source}. The download cost can be derived from the size of $Y_k$ and the value of recovery threshold. \cite{von2013modern} shows that interpolating any degree $k$ polynomial can be done through $O(k \log^2k \log \log k)$ operations. The decoding complexity is obtained by incorporating the fact that polynomial $h(f(z), g(z))$ is of degree $\hat{M}$.

\end{theorem}

\subsection{Example}

In this subsection, we shall introduce an example of applying our proposed scheme to the problem of matrix multiplication. To see the motivation of our construction, we shall first introduce the notion of bi-linear complexity of matrix multiplication \cite{smirnov2013bilinear, strassen1969gaussian}.
\begin{definition}For matrices ${A}=\left[A_{k, \ell}\right]_{k \in[m], \ell \in[p]}$ and ${B}=\left[B_{\ell, j}\right]_{\ell \in[p], j \in[n]}$. Suppose $AB=C=[C_{k,j}]$. Then the bi-linear complexity is defined as the minimum number of multiplications for calculating $C$ from $A$ and $B$, which we shall denote as $R(m, p, n)$. Any tensors $a \in \mathbb{F}_q^{R \times m \times p}$, $b \in \mathbb{F}_q^{R \times p \times n}$, and $c \in \mathbb{F}_q^{R \times m \times n}$ satisfying the conditions below are equivalent to the existence of an upper bound construction with rank $R$ for bi-linear complexity.

\begin{equation}
\begin{gathered}
\sum_{r=1}^R c_{r, k, j}(\underbrace{\sum_{k^{\prime}=1}^m \sum_{\ell^{\prime}=1}^p a_{r, k^{\prime}, \ell^{\prime}} A_{k^{\prime}, \ell^{\prime}}}_{=\bar{A}_r})(\underbrace{\sum_{\ell^{\prime}=1}^p \sum_{j^{\prime}=1}^n b_{r, \ell^{\prime}, j^{\prime}} B_{\ell^{\prime}, j^{\prime}}}_{=\bar{B}_r}) \\
=\sum_{\ell=1}^p A_{k, \ell} B_{\ell, j}=C_{k, j}, \quad \forall k \in[m], j \in[n] .
\end{gathered}   
\end{equation}

\end{definition}
The use of bi-linear complexity allows for the transformation of the matrix multiplication problem ${C}={A B}$ into the computation of the products of two sets of matrices $\{\bar{A}_1,\cdots, \bar{A}_R \}$ and $\{\bar{B}_1,\cdots, \bar{B}_R \}$.


Suppose there are $2$ sources $S_1$ and $S_2$, each $S_i$ holding a secret data $W_i$. Denote $W=[W_1 W_2]$. Assume the user also has secret data $U$. The goal is to compute the matrix product $WU$ with the help of $N$ worker nodes. 

We shall partition $W$ and $U$ as follows:

\begin{equation}
W_1 = \begin{bmatrix}
      W_{1,1}  \\
      W_{2,1} \end{bmatrix}, W_2 = \begin{bmatrix}
      W_{1,2}  \\
      W_{2,2} \end{bmatrix}, U = \begin{bmatrix}
      U_{1,1} & U_{1,2}\\
      U_{2,1} & U_{2,2} \end{bmatrix}.   
\end{equation}

Strassen \cite{strassen1969gaussian} gives a construction of bi-linear
complexity $R = 7$ as follows:

\begin{equation}
\begin{array}{ll}
\bar{W}_1={W}_{1,1}+{W}_{2,2}, & \bar{U}_1={U}_{1,1}+{U}_{2,2} \\
\bar{W}_2={W}_{2,1}+{W}_{2,2}, & \bar{U}_2={U}_{1,1} \\
\bar{W}_3={W}_{1,1}, & \bar{U}_3={U}_{1,2}-{U}_{2,2} \\
\bar{W}_4={W}_{2,2}, & \bar{U}_4={U}_{2,1}-{U}_{1,1} \\
\bar{W}_5={W}_{1,1}+{W}_{1,2}, & \bar{U}_5={U}_{2,2} \\
\bar{W}_6={W}_{2,1}-{W}_{1,1}, & \bar{U}_6={U}_{1,1}+{U}_{1,2} \\
\bar{W}_7={W}_{1,2}-{W}_{2,2}, & \bar{U}_7={U}_{2,1}+{U}_{2,2}.
\end{array}    
\end{equation}

Define $M_i=\bar{W}_i\bar{U}_i (i \in [7])$, then 
\begin{equation} \label{strassen}
 WU =  \begin{bmatrix}
      M_1 + M_4 - M_5 + M_7 & M_3 + M_5\\
      M_2 + M_4 & M_1 - M_2 + M_3 + M_6 \end{bmatrix}.  
\end{equation}

Assume the privacy protection level $X=2$. We will select any $9$ distinct elements $\beta_1, \cdots, \beta_9$ from $\mathbb{F}_q$. We then select $20$ distinct elements 
 $\{\alpha_i\}_{i \in [20]}$ from $\mathbb{F}_q$ such that $\{\alpha_i\}_{i \in[20]} \cap\{\beta_j\}_{j \in[7]}=\emptyset$. 

Define 

\begin{align}
& f^{(1)}(z) \nonumber\\
 = & W_{2,1} 
  ( \prod_{l \in[9] \backslash\{2\}} \frac{z-\beta_l}{\beta_2-\beta_l} + \prod_{l \in[9] \backslash\{6\}} \frac{z-\beta_l}{\beta_6-\beta_l}  )
   +W_{1,1} \nonumber\\
   &( \prod_{l \in[9] \backslash\{1\}} \frac{z-\beta_l}{\beta_1-\beta_l} + \prod_{l \in[9] \backslash\{3\}} \frac{z-\beta_l}{\beta_3-\beta_l} + \prod_{l \in[9] \backslash\{5\}} \frac{z-\beta_l}{\beta_5-\beta_l} \nonumber\\
   &- \prod_{l \in[9] \backslash\{6\}} \frac{z-\beta_l}{\beta_6-\beta_l}) + \sum^{2}_{j=1} P^{(1)}_j  \prod_{l \in[7]} \frac{z-\beta_l}{\beta_{7+j}-\beta_l}.    
\end{align}

\begin{align}
& f^{(2)}(z) \nonumber\\
 = & W_{1,2} ( \prod_{l \in[8] \backslash\{5\}} \frac{z-\beta_l}{\beta_5-\beta_l} + \prod_{l \in[8] \backslash\{7\}} \frac{z-\beta_l}{\beta_7-\beta_l} ) +
 W_{2,2} \nonumber\\
 &( \prod_{l \in[8] \backslash\{1\}} \frac{z-\beta_l}{\beta_1-\beta_l} + \prod_{l \in[8] \backslash\{2\}} \frac{z-\beta_l}{\beta_2-\beta_l}  + \prod_{l \in[8] \backslash\{4\}} \frac{z-\beta_l}{\beta_4-\beta_l} 
  \nonumber \\
 &- \prod_{l \in[8] \backslash\{7\}} \frac{z-\beta_l}{\beta_7-\beta_l}) + \sum^{2}_{j=1} P^{(2)}_j  \prod_{l \in[7]} \frac{z-\beta_l}{\beta_{7+j}-\beta_l}.    
\end{align}

\begin{equation}
g(z) =   \sum_{j \in[7]} \bar{U}_{j}  \prod_{l \in[8] \backslash\{j\}} \frac{z-\beta_l}{\beta_j-\beta_l} +  \sum^{2}_{j=1} Q_j  \prod_{l \in[7]} \frac{z-\beta_l}{\beta_{7+j}-\beta_l}.
\end{equation}

We shall denote $P_j=\sum^{2}_{i=1} P^{(i)}_j$ ($j \in [2]$).

\begin{equation}
f(z) =   \sum_{j \in[7]} \bar{W}_{j}  \prod_{l \in[8] \backslash\{j\}} \frac{z-\beta_l}{\beta_j-\beta_l}+  \sum^{2}_{j=1} P_j  \prod_{l \in[7]} \frac{z-\beta_l}{\beta_{7+j}-\beta_l}.
\end{equation}

Whenever each worker $k$ receives all the encoded matrices ${\bar W}^{(i)}_k=f^{(i)}(\alpha_k)$ and ${\bar U}_k=g(\alpha_k)$ from user and all sources, it shall compute $Y_k=({\bar W}^{(1)}_k+{\bar W}^{(2)}_k){\bar U}_k$ and return $Y_k$ to the user. Assume there exists one Byzantine worker. Note the degree of polynomial $f(z)g(z)$ is $16$. Using the RS code decoding algorithm, any $19$ workers' results will be sufficient to decode $f(z)g(z)$. Then the proposed protocol may resist $1$ straggler. Note for $i \in [7]$, $f(\beta_i)g(\beta_i)=M_i$. Thus the matrix product $WU$ can be successfully retrieved by equation \ref{strassen}.

\textbf{Remark}: The construction adapts to polynomial sharing \cite{zhu2021improved} schemes similarly. Grouping techniques \cite{zhu2022generalized} can also be performed easily.

\section{Proof of Privacy} \label{privacy proof}

Suppose $X$ workers in some subset $\mathcal{X}$ collude, denote the workers indexes in $\mathcal{X}$ as $k_j$ ($j=1,2,\cdots,X$).

\begin{lemma}[Generalized Cauchy Matrix\cite{DBLP:books/daglib/0068880}] \label{LCC lemma} Let $\alpha_1, \cdots, \alpha_X$ and $\beta_1, \cdots, \beta_X$ be (pairwise) distinct elements from a finite field $\mathbb{F}_q$. Denote $l_j(x)$ a Lagrange basis polynomial of degree $X-1$ defined as follows:
$$
l_j(z)=\prod_{l \in[X] \backslash\{j\}} \frac{z-\beta_l}{\beta_j-\beta_l}, \quad \forall j \in[X] .
$$

Then the following generalized Cauchy matrix is invertible over $F_q$.

\begin{equation*}
    \left[\begin{array}{cccc}
l_1\left(\alpha_1\right) &  l_2\left(\alpha_1\right) & \ldots &  l_X\left(\alpha_1\right) \\
l_1\left(\alpha_2\right) & l_2\left(\alpha_2\right) & \ldots & l_X\left(\alpha_2\right) \\
\vdots & \vdots & \ddots & \vdots \\
l_1\left(\alpha_X\right) & l_2\left(\alpha_X\right) & \ldots & l_X\left(\alpha_X\right)
\end{array}\right]_{X \times X}.
\end{equation*}

\end{lemma}

We shall first prove that $I(W; \widetilde{W}_{\mathcal{X}})=0$, and the equality $I(U; \widetilde{U}_{\mathcal{X}})=0$ follows similarly.

Denote ${\bar P}^{(i)}_{k_j}=\sum_{t=K+1}^{K+X} P^{(i)}_t  \prod_{l \in[K+X] \backslash\{t\}} \frac{\alpha_{k_j}-\beta_l}{\beta_t-\beta_l}$.

\begin{align}
& I\left(W; \widetilde{W}_{\mathcal{X}}\right) \nonumber\\
= & H(\widetilde{W}_{\mathcal{X}}) - H(\widetilde{W}_{\mathcal{X}} \mid W) \\
= & H(\{{\bar W}^{(i)}_{k_1} \}_{i \in [S]}, \cdots, \{{\bar W}^{(i)}_{k_{X}} \}_{i \in [S]}) \nonumber\\
& -H\left(\{{\bar W}^{(i)}_{k_1} \}_{i \in [S]}, \cdots, \{{\bar W}^{(i)}_{k_{X}} \}_{i \in [S]} \mid W \right) \\
\leq & \sum_{i \in [S]} \sum_{j \in [X]} H({\bar W}^{(i)}_{k_j} ) \nonumber\\
& -H\left(\{{\bar P}^{(1)}_{k_j} \}_{j \in [X]}, \cdots, \{{\bar P}^{(S)}_{k_{j}} \}_{j \in [X]}  \right) \\
= & SXab\frac{S}{K}\log q - \sum_{i \in [S]} H(\{{\bar P}^{(i)}_{k_{j}} \}_{j \in [X]}) \\
= & SXab\frac{S}{K}\log q - S(Xab\frac{S}{K}\log q) \nonumber\\
= & 0 \nonumber.
\end{align}

In the above derivation, equations $15$ and $16$ come from the definition of mutual information. The fourth equality is clear from the independence of sources. Equation $18$ follows immediately from the entropy
of a uniformly distributed random variable on a finite field $\mathbb{F}_q$ and lemma \ref{LCC lemma}. Inequality $17$ can be derived from the fact that joint entropy is bounded by the sum of respective entropies.

The privacy of data is guaranteed by the following inequality and the fact that mutual information $I\left(W, U; \widetilde{W}_{\mathcal{X}}, \widetilde{U}_{\mathcal{X}}\right)$ is non-negative.

\begin{align}
& I\left(W, U; \widetilde{W}_{\mathcal{X}}, \widetilde{U}_{\mathcal{X}}\right) \nonumber\\
= & I\left(W, U; \widetilde{W}_{\mathcal{X}}\right)+I\left(W, U; \widetilde{U}_{\mathcal{X}} \mid \widetilde{W}_{\mathcal{X}}\right) \\
= & H\left(\widetilde{W}_{\mathcal{X}}\right)-H\left(\widetilde{W}_{\mathcal{X}} \mid W, U\right) \nonumber\\
& +H\left(\widetilde{U}_{\mathcal{X}} \mid \widetilde{W}_{\mathcal{X}}\right)-H\left(\widetilde{U}_{\mathcal{X}} \mid \widetilde{W}_{\mathcal{X}}, W, U\right) \\
\leq & H\left(\widetilde{W}_{\mathcal{X}}\right)-H\left(\widetilde{W}_{\mathcal{X}} \mid W\right) \nonumber\\
& +H\left(\widetilde{U}_{\mathcal{X}}\right)-H\left(\widetilde{U}_{\mathcal{X}} \mid U\right) \\
= & I\left(W ; \widetilde{W}_{\mathcal{X}}\right)+I\left(U ; \widetilde{U}_{\mathcal{X}}\right) \\
= & 0 \nonumber.
\end{align}

In the above derivation, equation $19$ comes from the chain rule for mutual information. Equations $20$ and $22$ follow immediately from the definition of mutual information. Inequality $21$ can be derived from the non-increasing of information when conditioning and independence of data $W$ and $U$.

Interestingly, we can show that the shared data by sources (user) will form a $X$-private MDS storage system on $N$ worker nodes \cite{ DBLP:journals/tit/ZhuYTL22, DBLP:journals/tifs/RavivK19}. 

\begin{lemma}
Suppose $\Phi_i ( i \in [X])$ are $X$ i.i.d. random variables following the uniform distribution on a finite filed $\mathbb{F}_q$. Then $\sum^{X}_{i=1} \Phi_i$ also follows a uniform distribution on $\mathbb{F}_q$.
\end{lemma}

\textbf{Proof of lemma $4.2$}:
We shall prove that $\Phi_1 + \Phi_2$ follows a uniform distribution on $\mathbb{F}_q$. Then the lemma is valid from a direct induction.

For any $x \in \mathbb{F}_q$, 
$P(\Phi_1 + \Phi_2=x) = P( \bigcup_{y \in \mathbb{F}_q} \{ \Phi_1=y, \Phi_2=x-y \})
= \sum_{y \in \mathbb{F}_q} P(\Phi_1=y)P(\Phi_2=x-y) 
= \frac{1}{q}$.

From the above lemma, it is clear that the shared data by sources (user) can be seen as constructed by a secure Lagrange storage code \cite{DBLP:journals/tit/ZhuYTL22}. Thus forming a $X$-private MDS storage system.


\bibliographystyle{IEEEtran}
\bibliography{IEEEabrv,ref}

\vfill

\end{document}